\documentclass[aps,prl,twocolumn,showpacs,amsmath,amssymb,floatfix,superscriptaddress]{revtex4}
\usepackage{graphicx}
\usepackage{latexsym}
\usepackage{amsfonts}
\begin{document}
\title{Fluctuations in a diffusive medium with gain}
\date{\today}
\author{Stefano Lepri}
\email{stefano.lepri@isc.cnr.it}
\affiliation{Consiglio Nazionale delle Ricerche, Istituto dei Sistemi Complessi, 
via Madonna del Piano 10, I-50019 Sesto Fiorentino, Italy}
\begin{abstract}
We present a stochastic model for amplifying, diffusive media like, 
for instance, random lasers. Starting from a simple random-walk model, 
we derive a stochastic partial differential equation for the energy 
field with contains a multiplicative random-advection term yielding 
intermittency and power-law distributions of the field itself. 
Dimensional analysis indicate that such features are more likely 
to be observed for small enough samples and in lower spatial dimensions. 
\end{abstract}



\maketitle


Wave transport in disordered media can be described as
a multiple scattering process in which waves are randomly scattered 
by a large number of separate elements \cite{Sheng2006}. 
To first approximation this gives rise to a diffusion process. 
A particularly interesting situation arises when gain is added to a 
random material. In optics this is realized, 
for instance, in the form of a suspension of micro particles with added 
laser dye or by grinding a laser crystal. 
If the total gain becomes larger then the losses, fluctuations grow and 
these systems exhibit a lasing threshold \cite{Letokhov68}, 
yielding the so called random laser (see e.g. 
Refs.~~\cite{Wiersma2008,Cao2003} and the bibliography therein). 
The complexity generated by the interplay between gain and disorder
leads to intriguing connections with other fundamental problems 
like Anderson localization \cite{Pradhan1994} or the physics of 
glasses \cite{Angelani2006}.
Besides their fundamental interest, random lasers are likely to 
have a technological impact as low-cost light sources. 

Similar situations occur in other branches of physics like
neutron diffusion in fissile materials, or  
stochastic wave growth \cite{Robinson1995} and 
acceleration of plasma particles \cite{Park1995} 
As known, the competition between growth and propagation or 
diffusion is also a basic mechanism of population dynamics 
and theoretical ecology \cite{Mikhailov1992}. 

Diffusive random lasing has been observed experimentally in various 
active random media, including powders, laser dye suspensions and 
organic systems \cite{Zolin1986,Lawandy94,Cao99}. Theoretical descriptions 
often relies on the diffusive approximation either by reaction-diffusion 
type of equations~\cite{Letokhov68,Wiersma96} or, on a more microscopic level, 
by the master-equation approach~\cite{Florescu04-1}. 
Monte Carlo simulations play of course an important role
\cite{Berger97,Mujumdar2004a}. 

One of the salient experimental features of random laser emission
is its large statistical variability~\cite{Mujumdar2004}.
Indeed, already within the diffusive approximation, 
the addition of gain (and saturation) naturally
generates fat-tailed distributions ~\cite{Sharma2006}
that stems from rare long light paths \cite{Mujumdar2004a}.
This mechanism is of pure statistical origin and 
that does not require localization or interference~\cite{Mujumdar2004}. 
It was also proposed that such systems can
exhibit L\'evy type statistics in the distribution of 
intensities~\cite{Sharma2006,Wu2008,Zaitsev09} and crossovers among  
different statistics has been predicted \cite{Lepri2007}.
Remarkably, those prediction were very recently confirmed
experimentally \cite{Uppu2012} (see also Ref.\cite{Zhu2012}).

In this work we present first  a simple stochastic model that can be 
analytically solved and that yields large statistical fluctuations.
From it, we derive an equation for the energy field which contains
a stochastic multiplicative term whose strength controls such
anomalous fluctuations. The magnitude of this term introduces another
scale in the problem which thus provides a criterion for the 
observability of such intermittency in the field distribution.     


We choose to describe isotropic diffusion of light in terms of an ensemble of 
independent random walkers each carrying a given energy 
(number of photons). This may be visualized as an ensemble of  ``beams" 
propagating independently throughout the sample, each interacting 
with an underlying atomic population providing a gain mechanism
via stimulated emission. This procedure of attaching an energy to a 
random walk of photons is a common practice in Monte Carlo simulations 
of absorption in complex materials like e.g. tissues \cite{Prahl1989}. 
It has been also employed previously for random lasers 
\cite{Mujumdar2004a,Lepri2007}. 
under the implicit assumption that all the photons generated 
by amplification are diffused in the same direction at each 
scattering event.

Let us denote by $x$ and $E$ the walker position and energy.
We discuss the one-dimensional case in which the walker resides 
on a finite interval,  $0\le x\le L$. 
The dynamics is formulated as follows. 
A new walker is generated at a random position by a spontaneous emission event 
with a rate $\gamma$ and initial energy $E=\varepsilon$.
In terms of the underlying active media, $\gamma$ 
denotes the spontaneous emission rate of the single atom. The walker 
position $x$ changes to $x\pm a $ according to a standard random walk rule
on a lattice with spacing $a$. 
At the same time, the walker energy $E$ may increase by one unit 
due to the process of stimulated emission, $E\to E+\varepsilon$, 
with a rate $\Gamma(E)$. The simplest choice would be $\Gamma(E)=\gamma E$
or, to mimic saturation effects we may consider a 
gain of the form $\Gamma(E) = \gamma E/(1+E/E_{s})$.
 
The probability
$P_{i}(n)$ for the walker to be at 
$x=ia$ having an energy $E=n\varepsilon$ evolves according to 
a master equation, that can be solved by standard methods \cite{future}. 
Here, to simplify further, 
we work directly in the continuum limit and treat $x$ and $E$ as continuous 
variables, obeying the Langevin equations \cite{future}
\begin{equation}
\dot x = \sqrt{2D}\,\xi ,\quad \dot E = \Gamma(E) + \sqrt{\Gamma(E)} \, \eta
\label{langevin}
\end{equation}
where $\xi,\eta$ are $\delta$-correlated Gaussian variables with 
zero average and $\langle \xi ^2\rangle = \langle \eta ^2\rangle=1$
(from now on $\langle \ldots \rangle$ denotes an average over realizations
of the process). 
To keep things as simple as possible,  
bulk absorption or the possibility that diffusion is affected by 
the energy are neglected from scratch but can be easily included.


To demonstrate that this dynamics naturally generate power-law 
distributions of energies, we solve the associated Fokker-Planck equation 
(It\^o interpretation of Eq.~(\ref{langevin}))
\begin{equation}
\dot P = D\frac{\partial^2 P}{\partial x^2} - \frac{\partial }{\partial E}
\left(\Gamma P -\frac12 \frac{\partial \Gamma P}{\partial E}\right) 
\label{fpe}
\end{equation}
where $P(x,E,t)$ is the probability of finding a walker with 
energy $E$ at $x$. 
Four boundary conditions are necessary on the contour of the domain
$[0,L]\times [1,\infty)$. To account for complete absorption at the 
boundaries we let $P(0,E,t) = P(L,E,t) =0$. Moreover, 
$P(x,1,t)\equiv f(x,t)$ is determined as a solution of
\begin{equation}
\dot f = D f''  - \gamma f + \gamma 
\label{ebc}
\end{equation}
where we approximated $\Gamma(1)\approx \gamma$.
This condition may appear somehow unusual and is justified as follows: 
the number of particles with unit energy increases at rate $\gamma$ and are 
free to diffuse without increasing their energy. Their number diminishes 
by a term $-\gamma f$ because they gain energy by amplification.
Note that Eq.~(\ref{ebc}) can be derived exactly from the underlying
master equation for the discrete variables \cite{future}. 

The stationary solution of Eq.~(\ref{fpe}) can be found by separation of variables 
$P(x,E)=Q(x)W(E)$: 
yielding $Q \sim \sin (k x)$ with $k=m\pi/L$ ($m$ integer) being the separation constant 
(the wavenumber) that somehow couples diffusion and gain. 
The physical origin of such a coupling is that $E$ depends on the 
actual path length spent by the walker within the sample: longer paths acquire a
larger energy and the high-$E$ statistics ultimately depend on this
mechanism. 
The equations for $W$ may 
be integrated exactly, but for our purposes it suffices to solve
an approximated form where energy-diffusion term in Eq.~(\ref{fpe})
is neglected, 
$W \approx \exp\left(-D k^2 \int^E {\Gamma^{-1}(E')}{dE'}\right)/\Gamma(E)$.
Thus for the linear gain $W$ has a power-law tail while 
for the saturating case
\begin{equation}
W(E) \;\propto\; \frac{\exp(-Dk^2E/\gamma E_s)}{E^{1+Dk^2/\gamma}}
\end{equation}
(we have also assumed $E_s\gg E \gg 1$). The general solution 
is a sum over the allowed $k$s.
As a further approximation, we consider only the first Fourier mode 
$k=\pi/L$, which is mathematically justified 
noting that $Q$ is fixed by the stationary solution of Eq.~(\ref{ebc})
with $f(0)=f(L)=0$ is ($\beta\equiv\sqrt{\gamma/D}$)
\begin{equation}
f(x) = 1 + \frac{\sinh(\beta(x-L)) - \sinh(\beta x)}{\sinh(\beta L)}.
\label{bc}
\end{equation} 
In the critical region (to be defined below),  
$\beta \sim \pi/L$ so that $f$ (and thus $Q$) is indeed very close 
to the shape of the first Fourier mode itself. 
The result is that the distribution
of energies displays a parameter-dependent power-law 
\begin{equation}
P(x,E) \; \propto \; \sin\left(\frac{\pi x}{L}\right) \frac{\exp(-\alpha E/E_s)}{E^{1+\alpha}}
\label{powlaw}
\end{equation}
where $\gamma_c \equiv D(\pi/L)^2$ and $\alpha \equiv \gamma_c/\gamma$
which is exponentially cut-off at $E\sim E_s$.
The origin of the power-law (\ref{powlaw}) is traced back to 
very long paths which, although exponentially rare, acquire 
an exponentially large energy while diffusing throughout 
the sample \cite{Sharma2006,Lepri2007}. 
Note that $\gamma=\gamma_c$ correspond to the case of a Cauchy-like
tail $\alpha=1$, $P(x,E) \propto E^{-2}$, yielding a diverging
average value of the energy for $E_s\to \infty$. 
It is thus natural to identify this as the 
``laser threshold'' for the model: it will be shown below 
that this coincide with the usual criterion of gain overcoming 
the losses. 


Sofar we have described the system in terms of a single walker properties. 
Consider a population of $M$ walkers, and denote by
$x_i$ and $E_i$ their position and energies, each obeying the equation
of motion (\ref{langevin}). We thus
introduce the energy density field
\begin{equation}
\phi(x,t) = \sum_i^{M(t)} \, E_i \, \delta\left(x-x_i(t)\right).
\label{phi}
\end{equation}
The number $M(t)$ fluctuates in time due to the fact 
that walkers are created (at a rate $\gamma L$) and absorbed at the boundaries
(at a rate $D(\pi/L)^2 M$), so that $M \sim \gamma L^3/D$.
For $\gamma \sim \gamma_c \sim D/L^2$ (which is the regime of interest here)   
$M/L\sim 1$ i.e. there are a few walkers per unit length 
and fluctuations must be very relevant there. 
The problem can be thus recasted \cite{Dean1996} in term of a Langevin equation
for $\phi$ \cite{future}
\begin{equation}
\frac{\partial \phi}{\partial t}=
D\frac{\partial^2 \phi}{\partial x^2} +\Gamma (\phi) + 
\frac{\partial }{\partial x}(v\phi) + s
\label{spde}
\end{equation}
As done above, we neglected the $\sqrt{\Gamma}$ term in Eq.(\ref{langevin}) 
which also avoids the usual interpretation problems of the white-noise process.
The same $\Gamma$ as in the single-walker case is used, which is justified 
since the walkers are non-interacting. 
The additive, spatially uncorrelated, spontaneous emission noise $s(x,t)$ 
is a Poissonian process whereby $\phi$ is increased by $\varepsilon$ at random times, 
whose separation $\tau$ is a distributed as $\gamma \exp(-\gamma \tau)$.
The process $v(x,t)$ is a Gaussian-distributed with
\begin{equation}
 \langle v(x,t) v(x',t')\rangle = D \ell \Delta(x-x') \delta(t-t')
 \label{vcor}
\end{equation}
where the spatial scale $\ell$ is introduced for dimensional 
consistency. We further assume that this noise has a finite correlation 
in space, described by the smooth function $\Delta(x)$, an assumption 
often done in other contexts \cite{Kupiainen2007} 
to insure that the corresponding terms is well-defined.


The mean-field equation obtained by neglecting 
fluctuations in Eq.~(\ref{spde}) and replacing $s$ by 
its average $\gamma$ is a simplified, one-dimensional 
version of Letokhov equation for random lasers \cite{Letokhov68}. 
The steady state solution $\bar \phi(x)$ 
is not identically zero due to the $\gamma$ term. Neglecting this, 
we obtain that $\bar\phi$ destabilizes 
for $\gamma=\gamma_c$ thus defining a threshold as assumed above \cite{note}.

As usual for stochastic partial differential equation \cite{Garcia-Ojalvo1999}
Eq.~(\ref{spde}) is intended as a limit of some discretization on a finite mesh 
whose spacing we denote by $\Delta x$. 
For definiteness and for actual numerical investigation, we choose 
the discretized equation for $\phi_i$ to be ($i=0,\ldots,N+1$, 
$(N+1)\Delta x = L$)
\begin{eqnarray}
\label{discrete}
\dot \phi_i &=& 
D \left[ \phi_{i+1} + \phi_{i-1}  
-2 \phi_{i}\right] +\Gamma (\phi_i) \\
&+&   \frac{1}{2} (v_{i+1}\phi_{i+1} - v_{i-1}\phi_{i-1}) + s_i
\nonumber
\end{eqnarray} 
with $v_i$ being independent Gaussian variables with $\langle v_i\rangle=0$ 
and $\langle v_i^2\rangle=D\ell$ (we set $\Delta x=1$). 
Such discretization is legitimate since $\Delta(x)$ is assumed to
be smooth over distances of the order of the mean free path.
The boundary conditions are $\phi_0=\phi_{N+1}=0$
and we also impose $v_1=v_N=0$ to ensure that the multiplicative 
term conserves the total energy $\sum_i\phi_i$.
For the time derivative, we use a simple Euler discretization 
with a time step $\Delta t$. 

In the simulations, we choose $s_i$ to be either a Poisson process, 
(i.e. $s_i=1$ or zero with probability $\gamma\Delta t$ and $1-\gamma\Delta t$,
respectively) or a Gaussian with the same average, 
$s_i= \gamma + \sqrt{\gamma}\, r_i$, $r_i$ being normally-distributed and 
independent random numbers with $\langle r_i\rangle=0$ 
and $\langle r_i^2\rangle=1$. Although the first choice is closer to the original 
formulation of the model, we found that in practice, the two processes 
yield almost indistinguishable results. 

\begin{figure}[ht]
\begin{center}
\includegraphics[width=\columnwidth,clip]{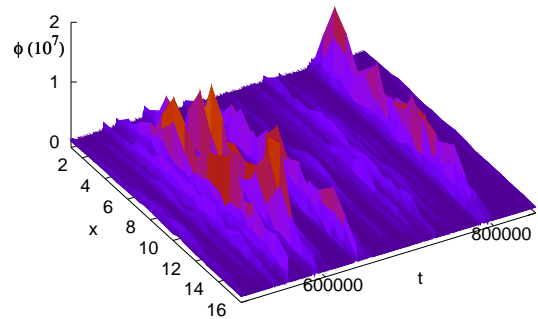}
\caption{(Color online) Intermittent evolution of the intensity $\phi(x,t)$ obtained 
by numerical integration of Eq.~(\ref{discrete});
$L=16$, $\gamma=1.02\gamma_c$, $D=1/2$, $\ell=1$, 
$\Gamma(\phi)=\gamma \phi/(1+\phi/\phi_s)$, $\phi_s=10^8$; 
here and in the following $\Delta t = 0.01$; $s_i$ is Poissonian (see text).}
\label{fig.1}
\end{center}
\end{figure}

The term in $v$ of Eq.(\ref{spde}) is the leading 
stochastic correction to the mean-field evolution. It
is a multiplicative process and can be regarded as  
a kind of random advection whereby fluctuations are transported 
almost coherently, while keeping the total energy conserved. It thus  
acts against the diffusive term that tends to smooth 
out fluctuations. As a result, the field is highly 
intermittent in time, with large-amplitude bursts
emerging from a lower-amplitude background (Fig.~\ref{fig.1}). It is thus 
intuitively plausible that such term is responsible of most of the 
nontrivial statistics of the field. As a matter of fact, 
random advection dynamics is known to yield 
strongly non-Gaussian distributions, and even power-law tails \cite{Takayasu1993}. 
Actually, equations similar to (\ref{discrete}) 
(with linear gain and periodic boundary conditions) 
have been studied in Refs.~\cite{Deutsch1993,Deutsch1994} 
to model an active scalar (e.g. a temperature field) 
convected by a random velocity field. It was argued 
that power law tails generically
arise when, as here, both multiplicative and additive noises are present. 
Actually, there are two 
sources of additive noise. One is the spontaneous 
emission term $s$; the other stems from the fact that the 
deterministic, steady state value is $\bar\phi \neq 0$, 
thus yielding a additive contribution of order 
$\partial (v\bar\phi)/\partial x$.
There is however a crucial difference, dictated
by the physical origin of the noise: 
the advective term is a finite-size effect which decreases 
with the system size. This can be demonstrated by dimensional analysis.
To be more general, let us consider the extension of Eq.~(\ref{spde})
to $d$ dimensions, 
\begin{equation}
\frac{\partial \phi}{\partial t}=D\nabla^2 \phi +\gamma \phi  + 
\nabla \cdot (\mathbf{v}\phi) + \gamma
\label{ddim}
\end{equation}
where $\mathbf{v}$ is a $d-$dimensional vector whose components
$v_\nu$ ($\nu=1,\ldots, d$) are Gaussian distributed and 
satisfy a relation akin to Eq.~(\ref{vcor}), 
\begin{equation}
 \langle v_\nu(\mathbf{x},t) v_{\nu'}(\mathbf{x}',t')\rangle = D \ell^d\, 
 \delta_{\nu,\nu'}\,\Delta_d(\mathbf{x}-\mathbf{x}') \delta(t-t'),
 \label{vcord}
\end{equation}
with $\Delta_d$ being a generalization of the function in Eq.~(\ref{vcor}).
For simplicity, we neglected again the $\sqrt{\Gamma}$ term, replaced
$s$ by its average $\gamma$ and restricted to the simpler case 
$\Gamma(\phi) = \gamma \phi$. Let us rewrite Eq.~(\ref{spde}) introducing the 
dimensionless variables $\mathbf{x}/L \to \mathbf{x}$, $Dt/L^2 \to t$, $\phi/\varepsilon \to \phi$ 
and we also rescale $D\gamma/L^2  \to \gamma$. 
The noise rescales accordingly as $\sqrt{L^{d+2}/D}\,\mathbf{v} \to \mathbf{v}$.
Thus, Eq. (\ref{spde}) is  written in the dimensionless form 
\begin{equation}
\frac{\partial \phi}{\partial t}=\nabla^2 \phi +\gamma \phi  + 
\lambda(L) \nabla \cdot (\mathbf{u}\phi) + \gamma
\end{equation}
where each component of $\mathbf{u}$ is again an uncorrelated Gaussian variable
with $\langle u_\nu^2\rangle=1$ and we have made explicit the rescaled noise strength 
which turns out to be $\lambda \propto (\ell/L)^\frac{d}{2}$. As a 
consequence, we expect that the multiplicative noise to yield
sizeable power-law tails only for small enough sizes.
When $L\to\infty$, the diffusive term in Eq.~(\ref{ddim}) dominates 
and power law are no longer observable \cite{Deutsch1994}.
 
\begin{figure}[ht]
\begin{center}
\includegraphics[width=\columnwidth,clip]{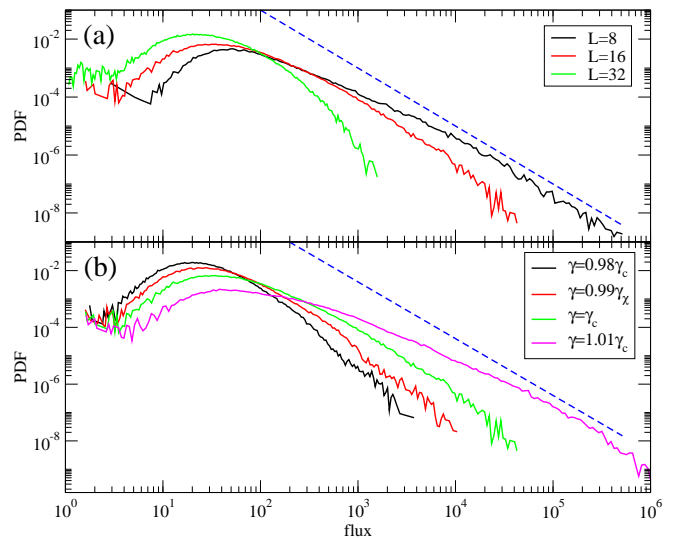}
\caption{(Color online) Probability distribution functions of the flux 
$J$ for (a) $\gamma=\gamma_c$ and increasing $L$ values (right to left);
(b) $L=16$, increasing $\gamma$ values (left to right). 
The straight dashed line corresponds to the power 
law $\alpha=1$ expected from Eq.~(\ref{powlaw}). 
Flux data have been binned over consecutive time windows 
of duration $T_W=10$, over a run of about $10^6$ time units. 
Additive noise is Gaussian (see text); other parameters 
as in previous figure.}
\label{fig.2}
\end{center}
\end{figure}

The numerical simulations of Eq.~(\ref{discrete}) are in qualitative 
agreement with the above argument. We monitored the distribution
of the field at the center of the mesh $\phi_{N/2}$ as well 
as the the outcoming flux $J=D[\phi_1(t) + \phi_N(t)]/2$ as function 
of time. 
This quantity is of experimental interest, being related to the emission
spectra.  The tail of the distributions  of both observables display, upon
increasing $L$, a smooth crossover  from a power-law tail with an exponent
$\alpha$ very close to the one  predicted by Eq.~(\ref{powlaw}) to a faster
decay  (see Fig.\ref{fig.2}a). The same occurs for fixed $L$ upon approaching
the threshold  $\gamma_c$ (Fig.\ref{fig.2}b).

\begin{figure}[ht]
\begin{center}
\includegraphics[width=\columnwidth,clip]{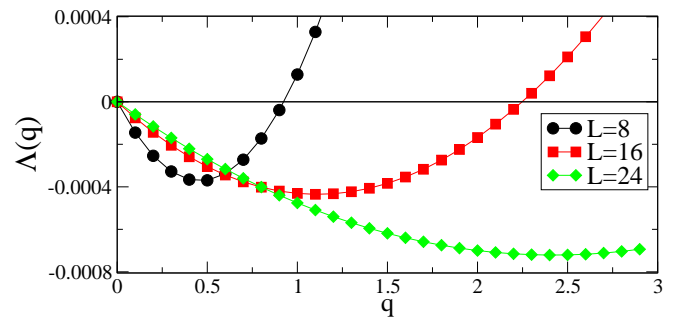}
\caption{(Color online) 
The generalized finite-time Lyapunov exponent $\Lambda(q)$ for $\tau=200$
and different system lengths $L$. 
}
\label{fig.3}
\end{center}
\end{figure}

The existence of fat tails is intimately related to the the possibility
for a spontaneous fluctuation to grow well beyond the average.
The indicator to quantify this is the generalized Lyapunov exponent 
\cite{Benzi1985}. For a perturbation $\delta \phi(x,t)$ of the 
field, which evolves according to the linearized version of 
Eq.(\ref{spde}), let $R(\tau) = \|\delta \phi(t+\tau)\|/\|\delta \phi(t)\|$
be the response function after a time $\tau$ to a disturbance at time $t$.
The generalized finite-time Lyapunov exponent $\Lambda(q)$ is then defined 
by $\overline{R^q(\tau)}\sim\exp(\Lambda(q)\tau)$ 
where the overline denote a time average.
If $\Lambda(q) > 0$ for large enough $q$ then the system has a finite
probability that a small perturbation results in a large one.
Moreover, the deviations of $\Lambda(q)$ from a linear behaviour
in $q$ are a measure of intermittency \cite{Benzi1985}.
Fig.\ref{fig.3} shows that $\Lambda>0$ for $q>q_*$ with $q_*$ increasing 
with $L$. This signals that 
events far from the average become increasingly rare, in 
agreement with the above dimensional arguments.
Also, for $L=8$ 
$q_*=0.92$ which, according to Ref.~\cite{Deutsch1993} 
would yield a power-law decay with an exponent $1+q_*=1.92$, 
in agreement with data in Fig.~\ref{fig.2}.

To summarise, we have presented a simple model for diffusive random media 
with gain which yields power law distribution of the intensities. 
Our main result is the Langevin equation for the energy density 
field, Eq.(\ref{spde}), that establishes a novel connection
between the physics of scattering media with gain (e.g. random lasers)
and the theory of nonequilibrium phenomena in spatially extended systems.
The random advection, multiplicative term in Eq.(\ref{spde}) competes 
with diffusive
and gain terms and is responsible for unconventional fluctuations. 
Its relevance is gauged by the effective noise strength $\lambda(L)$.
This means that L\'evy-like fluctuations are more likely to be 
observed in lower dimensions (for instance in $d=2$) and when the 
mean free path of photons is not too short with respect to the sample
size. In other words, as in the case of advective mixing \cite{Deutsch1993},
small systems are also necessary to observe the effect. 
For an experimental validation of the model for random lasers, 
one should compare the emission statistics for samples with 
different $\lambda(L)$. This is achieved, for instance, by changing the 
gain volume (controlled by pumping) and/or the mean-free-path 
which can be tuned by varying the density of scatterers. 


\acknowledgments

I thank S. Cavalieri, F. Ginelli and R. Livi.


\begin{thebibliography}{31}
\bibitem{note} Actually, the noise term may shift such threshold 
by some amount proportional to the noise strength \cite{Robinson1997}, but this 
issue is not relevant for the present discussion.
\end{thebibliography}
\end{document}